\documentclass[english]{scrartcl}
\usepackage[utf8]{inputenc}
\DeclareUnicodeCharacter{A0}{~}
\usepackage[T1]{fontenc}
\usepackage[style=numeric-comp,backend=bibtex]{biblatex}
\bibliography{wifi}
\usepackage[left=1in,right=1in]{geometry}
\usepackage[hang,small,bf]{caption}
\usepackage[autostyle,english=british]{csquotes}
\usepackage[british]{babel}
\usepackage{varioref}
\usepackage{hyperref}
\hypersetup{colorlinks=true, linkcolor=black, citecolor=black,
  filecolor=black, urlcolor=black, bookmarksdepth=subsection,
  unicode=true}
\usepackage{graphicx}
\usepackage{wrapfig}
\usepackage{float}
\usepackage{longtable}
\usepackage{multirow}
\usepackage{tabularx}
\usepackage{tabu}
\usepackage{framed}
\usepackage{booktabs}
\usepackage{tikz}
\usepackage{algorithm}
\usepackage[noend]{algpseudocode}
\usetikzlibrary{arrows,positioning,intersections,shadows,shapes}
\usepackage{scrpage2}
\clearscrheadings
\usepackage{usenix}
\usepackage{amsmath}
\usepackage{bbm}
\usepackage{listings}
\usepackage{url}
\usepackage{subcaption}
\usepackage{upgreek}
\usepackage{siunitx}

\author{
{\upshape Toke Høiland-Jørgensen}\\
Karlstad University
\and
{\upshape Michał Kazior}\\
Tieto Poland
\and
{\upshape Dave Täht}\\
TekLibre
\and
{\upshape Per Hurtig}\\
Karlstad University
\and
{\upshape Anna Brunstrom}\\
Karlstad University}
\date{}
\title{Ending the Anomaly: Achieving Low Latency and Airtime Fairness in WiFi}
\hypersetup{
 pdfauthor={Toke Høiland-Jørgensen, Michał Kazior, Dave Täht, Per Hurtig and Anna Brunstrom},
 pdftitle={Ending the Anomaly: Achieving Low Latency and Airtime Fairness in WiFi},
 pdfkeywords={},
 pdfsubject={},
 pdfcreator={Emacs 26.0.50.2 (Org mode 8.3.6)},
 pdflang={English}}
\begin{document}

\maketitle
\begin{abstract}
With more devices connected, delays and jitter at the WiFi hop become more
prevalent, and correct functioning during network congestion becomes more
important. However, two important performance issues prevent modern WiFi from
reaching its potential: Increased latency under load caused by excessive
queueing (i.e. \emph{bufferbloat}) and the 802.11 performance anomaly.

To remedy these issues, we present a novel two-part solution: We design a new
queueing scheme that eliminates bufferbloat in the wireless setting. Leveraging
this queueing scheme, we then design an airtime fairness scheduler that operates
at the access point and doesn't require any changes to clients.

We evaluate our solution using both a theoretical model and experiments in a
testbed environment, formulating a suitable analytical model in the process. We
show that our solution achieves an order of magnitude reduction in latency under
load, large improvements in multi-station throughput, and nearly perfect airtime
fairness for both TCP and downstream UDP traffic. Further experiments with
application traffic confirm that the solution provides significant performance
gains for real-world traffic.We develop a production quality implementation of
our solution in the Linux kernel, the platform powering most access points
outside of the managed enterprise setting. The implementation has been accepted
into the mainline kernel distribution, making it available for deployment on
billions of devices running Linux today.
\end{abstract}

\section{Introduction}
\label{sec:org0d29e57}
As more mobile devices connect to the internet, and internet connections
increase in capacity, WiFi is increasingly the bottleneck for users of the
internet. This means that congestion at the WiFi hop becomes more common, which
in turn increases the potential for bufferbloat at the WiFi link, severely
degrading performance \cite{good-bad-wifi}.

The 802.11 performance anomaly \cite{heusse_performance_2003} also
negatively affects the performance of WiFi bottleneck links. This is a
well-known property of WiFi networks: if devices on the network operate at
different rates, the MAC protocol will ensure \emph{throughput fairness} between
them, meaning that all stations will effectively transmit at the lowest rate.
The anomaly was first described in 2003, and several mitigation strategies have
been proposed in the literature
(e.g., \cite{tan_time-based_2004,joshi_airtime_2008}), so one would expect the
problem to be solved. However, none of the proposed solutions have seen
widespread real-world deployment.

Recognising that the solutions to these two problems are complementary, we
design a novel queue management scheme that innovates upon previous solutions to
the bufferbloat problem by adapting it to support the 802.11n protocol. With
this queueing structure in place, eliminating the performance anomaly becomes
possible by scheduling the queues appropriately. We develop a deficit-based
airtime fairness scheduler to achieve this.

We implement our solution in the WiFi stack of the Linux kernel. Linux is
perhaps the most widespread platform for commercial off-the-shelf routers and
access points outside the managed enterprise, and hundreds of millions of users
connect to the internet through a Linux-based gateway or access point on a daily
basis. Thus, while our solution is generally applicable to any platform that
needs to support 802.11n, using Linux as our example platform makes it possible
to validate that our solution is of production quality, and in addition gives
valuable insights into the practical difficulties of implementing these concepts
in a real system.

The rest of this paper describes our solution in detail, and is structured as
follows: Section \ref{sec:background} describes the bufferbloat problem in the
context of WiFi and on the WiFi performance anomaly, and shows the potential
performance improvement from resolving them. Section \ref{sec:solution}
describes our proposed solution in detail and Section \ref{sec:evaluation}
presents our experimental evaluation. Finally, Section \ref{sec:related-work}
summarises related work and Section \ref{sec:conclusion} concludes.

\section{Background}
\label{sec:background}
In this section we describe the two performance issues we are trying to solve:
Bufferbloat in the WiFi stack and the 802.11 performance anomaly. We explain why
these matter, and show the potential benefits from solving them.

\subsection{Bufferbloat in the context of WiFi}
\label{sec:org4a60b30}
Previous work on eliminating bufferbloat has shown that the default buffer
sizing in many devices causes large delays and degrades performance. It also
shows that this can be rectified by introducing modern queue management to the
bottleneck link \cite{good-bad-wifi,cablelabs,much-ado}. However, this does not
work as well for WiFi; prior work has shown that neither decreasing buffer sizes
\cite{showail_buffer_2016} nor applying queue management algorithms to the WiFi
interface \cite{good-bad-wifi} can provide the same reduction in latency under
load as for wired links.

\begin{figure}[htbp]
\centering
\includegraphics[width=\linewidth]{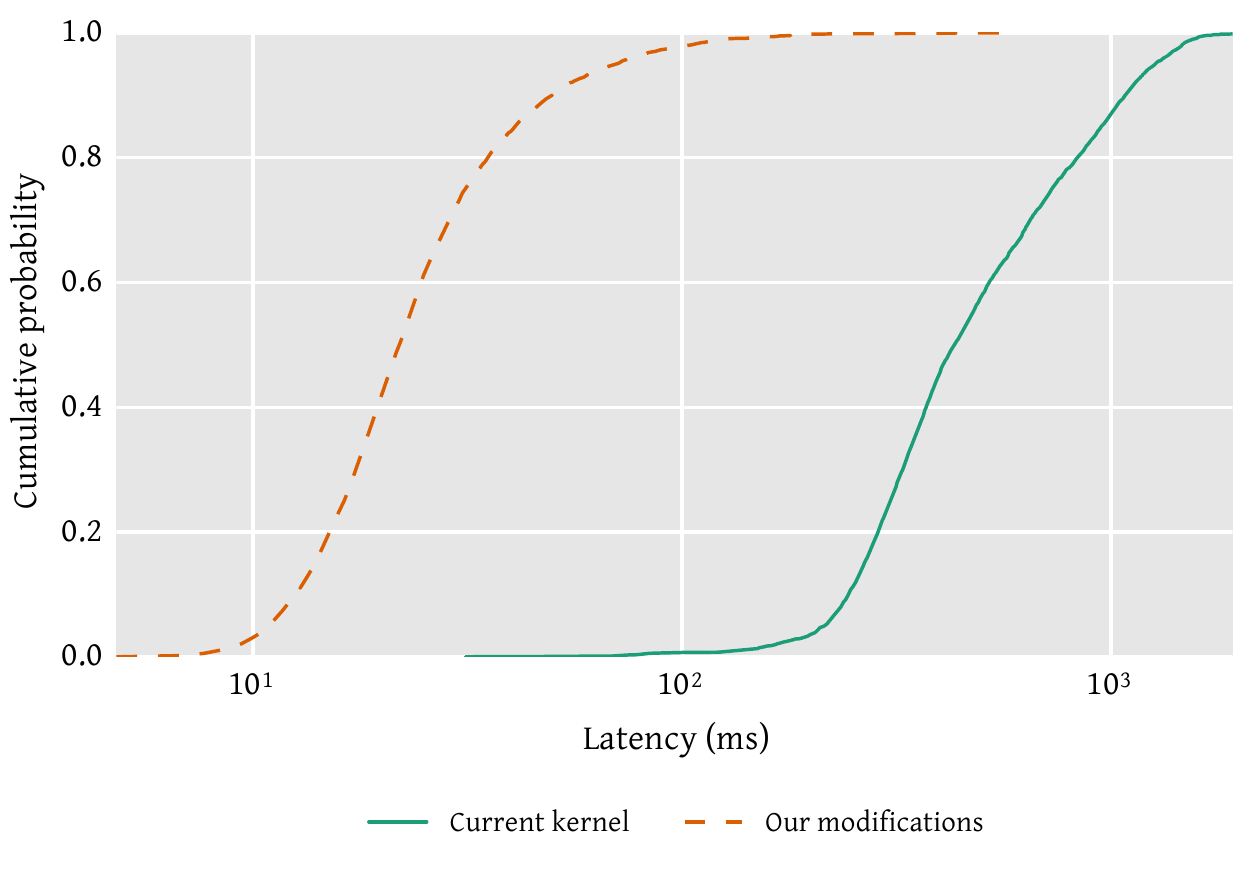}
\caption{\label{fig:tcp-up-latency-partial}
Latency of an ICMP ping flow with simultaneous TCP download traffic, before and after our modifications.}
\end{figure}

Figure \ref{fig:tcp-up-latency-partial} showcases the potential gain from fixing
bufferbloat in WiFi. The figure shows a latency measurement (ICMP ping)
performed simultaneously with a simple TCP download to each of the stations on
the network. The solid line shows the state of the Linux kernel in its default
configuration: Several hundred milliseconds of added latency. The dashed line
shows the effects of applying the solution we propose in this paper: A latency
reduction of an order of magnitude.

\begin{figure}[htbp]
\centering
\includegraphics[width=\linewidth]{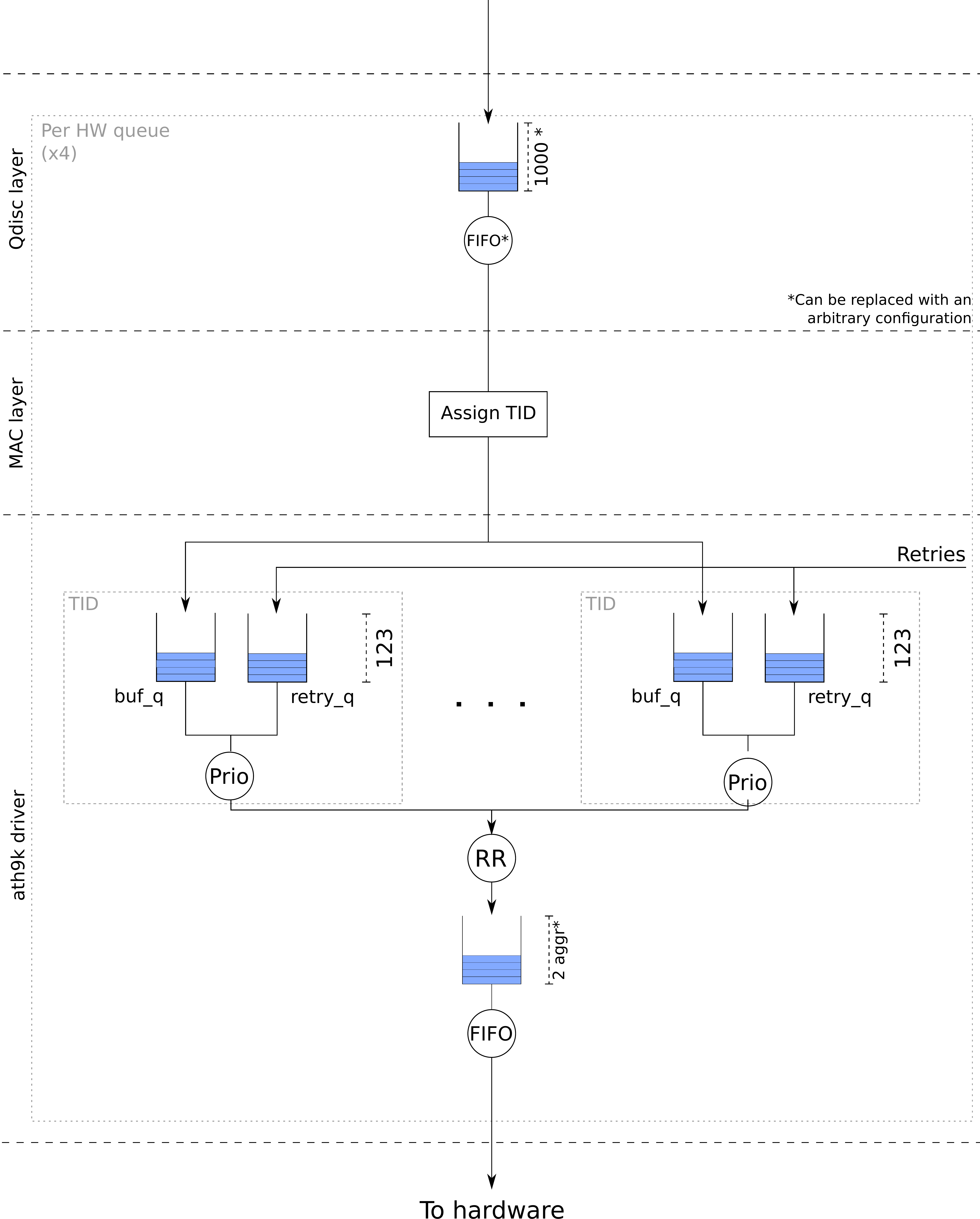}
\caption{\label{fig:linux-queue-pre}
The queueing structure of the Linux WiFi stack.}
\end{figure}

The reason for the limited effect of prior solutions is queueing in the lower
layers of the wireless network stack. For Linux, this is clearly seen in the
queueing structure, depicted in Figure \ref{fig:linux-queue-pre}. The upper queue
discipline ("qdisc") layer, which is where the advanced queue management schemes
can be installed, sits above both the mac80211 subsystem (which implements the
base 802.11 protocol) and the driver. As the diagram shows, there is significant
unmanaged queueing in these lower layers, limiting the efficacy of the queue
management schemes and leading to increased delay. Because queueing is needed at
a low layer in order to build aggregates (more on this later), an integrated
queueing scheme is needed to bring the benefits of modern queue management to
WiFi.

\subsection{Airtime fairness}
\label{sec:org2b86a95}
The 802.11 performance anomaly was first described for the 802.11b standard
in \cite{heusse_performance_2003}, which showed that in a wireless network with
differing rates, each station would achieve the same effective throughput even
when their rates were different. Later work has showed both analytically and
experimentally that time-based fairness improves the aggregate performance of
the network \cite{tan_time-based_2004}, and that the traditional notion of
proportional fairness \cite{laddomada_throughput_2010} translates to airtime
fairness when applied to a WiFi network \cite{jiang_proportional_2005}.

This latter point is an important part of why airtime fairness is desirable:
Proportional fairness strikes a balance between network efficiency and allowing
all users a minimal level of service. Since a wireless network operates over a
shared medium (the airwaves), access to this medium is the scarce resource that
needs to be regulated. Achieving airtime fairness also has the desirable
property that it makes a station's performance dependent on the \emph{number} of
active stations in the network, and not on the performance of each of those
other stations.

To quantify the expected gains of airtime fairness in the context of 802.11n,
the following section develops an analytical model to predict throughput and
airtime usage.

\subsubsection{An analytical model for 802.11n}
\label{sec:orgdfe3ac0}
The models in \cite{heusse_performance_2003} and \cite{tan_time-based_2004} give
analytical expressions for expected throughput and airtime share for 802.11b
(the latter also under the assumption of airtime fairness). Later
work \cite{kim_adaptive_2015} updates this by developing analytical expressions
for packet sizes and transmission times for a single station using 802.11n.
However, this work does not provide expressions for predicting throughput and
airtime usage. In this section we expand on the work of \cite{kim_adaptive_2015}
to provide such an expression.

The 802.11n standard permits two types of aggregation: One is A-MPDU
aggregation, which works by assigning a MAC-level header and a Frame Check
Sequence (FCS) to each data packet, turning it into a MAC-level Protocol Data
Unit (MPDU). Several MPDUs can be aggregated into an A-MPDU, which then has a
frame header prepended and is transmitted in a single operation. The other type
of aggregation, called A-MSDU, combines several packets in a single MPDU, but to
simplify the model we exclude that in the following treatment.\footnote{For the version that includes A-MSDU aggregation, see
\cite{kim_adaptive_2015}.}

For the following exposition, we assume a set of active stations, \(I\). Each
station, \(i\), transmits aggregates consisting of \(n_i\) packets of size \(l_i\)
bytes. Furthermore, we assume that all packets in an aggregate are of the same
length. All constants are defined by the standard, and listed
in \cite{kim_adaptive_2015}. The size of such an aggregate then becomes:

\begin{equation}
L(n_i,l_i) = n_i(l_i+L_{delim}+L_{mac}+L_{FCS}+L_{pad})
\end{equation}

where \(L_{delim}=4\) is the MPDU delimiter, \(L_{mac}=34\) is the MAC header,
\(L_{FCS}=4\) is the frame check sequence and \(L_{pad}\) is the padding required to
get the total length to a multiple of four bytes.

From this we can derive the transmission time of the data portion of a packet
transmitted at the PHY rate \(r_i\) (measured in bps):

\begin{equation}
T_{data}(n_i,l_i,r_i) = T_{phy}+\frac{8L(n_i,l_i)}{r_i}
\end{equation}

where \(T_{phy}=32\,\mu s\) is the time required to transmit the PHY header.

From this we can compute the expected rate to the station assuming no errors or
collisions:

\begin{equation}
R(n_i,l_i,r_i) = \frac{n_i l_i}{T_{data}(n_i,l_i,r_i)+T_{oh}}
\end{equation}

where \(T_{oh}=T_{DIFS}+T_{SIFS}+T_{ack}+T_{BO}\) is the per-transmission
overhead, which consists of the Distributed Inter-Frame Space, \(T_{DIFS}=34\,\mu
s\), the Short Inter-Frame Space, \(T_{SIFS}=16\,\mu s\), the average block
acknowledgement time, \(T_{ack}\), and the average back-off time before
transmission, \(T_{BO}\). The latter two values are discussed extensively in
\cite{arif_throughput_2014} (along with cases where various forms of transmission
errors occur). However, following \cite{kim_adaptive_2015}, we limit ourselves to
estimating \(T_{BO}\simeq T_{slot}(CW_{min}/2)=68\,\mu s\) in the case of no
collisions and \(T_{ack} = T_{SIFS} + 8*58 / r_i\).

Turning to airtime fairness, we borrow two insights from the analysis in
\cite{tan_time-based_2004}:

\begin{enumerate}
\item The rate achieved by station \(i\) is simply given by the baseline rate it can
achieve when no other stations are present (i.e., \(R(n_i,l_i,r_i)\))
multiplied by the share of airtime available to the station.

\item When airtime fairness is enforced, the available airtime is divided equally
among the stations (by assumption). When it is not, the airtime share
available to station \(i\) is the ratio between the time that station spends on
a single transmission (i.e., \(T_{data}(n_i,l_i,r_i)\)) and total time all
stations spend doing a single transmission each.
\end{enumerate}

With these points in mind, we express the expected airtime share \(T(i)\) and rate
\(R(i)\) as:

\begin{align}
T(i) &=
  \begin{cases}
    \frac{1}{|I|} & \quad \text{with fairness} \\
    \frac{T_{data}(n_i,l_i,r_i)}{\sum_{j\in I} T_{data}(n_j,l_j,r_j)} & \quad \text{otherwise} \\
  \end{cases} \\
R(i) &= T(i) R(n_i,l_i,r_i)
\end{align}

Using the above, we can calculate the expected airtime share and effective rate
for each station in our experimental setup, assuming a fixed packet size of
\(1500\) bytes. The assumption of no contention holds because all data is
transmitted from the access point. As the queueing structure affects the
achievable aggregation level (and thus the predictions of the model), we use the
measured average aggregation levels in our experiments as input to the model.

\begin{table}[htb]
\centering
\small
\begin{tabular}{lcrrrr}
\toprule
\multirow{2}{6mm}{Aggr size} & $T(i)$ & \multicolumn{4}{c}{Rates (Mbps)} \\ \cmidrule{3-6}
 &  & PHY & Base & $R(i)$ & Exp\\
\midrule
\multicolumn{6}{c}{Baseline (FIFO queue)\footnotemark[2]} \\
$4.47$ &  $10\%$ & $144.4$ & $97.3$ & $9.7$ & $7.1$ \\
$5.08$ &  $11\%$ & $144.4$ & $101.1$ & $11.4$ & $6.3$ \\
$1.89$ &  $79\%$ & $7.2$ & $6.5$ & $5.1$    & $5.3$\\ \cmidrule{5-6}
\multicolumn{4}{l}{\textbf{Total}} & $26.4$ & $18.7$ \\
\midrule
\multicolumn{6}{c}{Airtime Fairness} \\
$18.44$ &  $33\%$ & $144.4$ & $126.7$ & $42.2$ & $38.8$\\
$18.52$ &  $33\%$ & $144.4$ & $126.8$ & $42.3$ & $35.6$\\
$1.89$ &  $33\%$ & $7.2$ & $6.5$ & $2.2$       & $2.0$\\  \cmidrule{5-6}
\multicolumn{4}{l}{\textbf{Total}} & $86.8$ & $76.4$ \\
\bottomrule
\end{tabular}
\caption{\label{tbl:airtime-rates}\small
Calculated airtime, calculated rate and measured rate for the three stations (two fast
and one slow) in our experimental setup. The aggregation size is the measured
mean aggregation size (in packets) from our experiments and the measured rates
(Exp column) are mean UDP throughput values.}
\end{table}

The model predictions, along with the actual measured throughput in our
experiments, are shown in Table \ref{tbl:airtime-rates}. The values will be
discussed in more detail in Section \ref{sec:evaluation}, so for now we will
just remark that this clearly shows the potential of eliminating the performance
anomaly: An increase in total throughput by up to a factor of five.

\section{Our solution}
\label{sec:solution}
We focus on the access point scenario in formulating our solution, since a
solution that only requires modifying the access point makes deployment easier
as there are fewer devices to upgrade. However, WiFi client devices can also
benefit from the proposed queueing structure. The rest of this section describes
the two parts of our solution in detail, and outlines the current implementation
status in Linux.

\footnotetext[2]{The aggregation size and throughput values vary quite a bit for this test, because of the randomness of the FIFO queue emptying and filling. We use the median value over all repetitions of the per-test mean throughput and aggregation size; see the online appendix for graphs with error bars.}
\addtocounter{footnote}{1}

\subsection{A bloat-free queueing structure for 802.11}
\label{sec:org455e1f5}
An operating system networking stack has many layers of intermediate queueing
between different subsystems, each of which can add latency. For specialised
systems, it is possible to remove these queues entirely, which achieves
significant latency reductions \cite{adam_belay_ix:_2014}. While such a radical
restructuring of the operating system is not always possible, the general
principle of collapsing multiple layers of queues can be applied to the problem
of reducing bufferbloat in WiFi. What we propose here is such an integrated
queueing structure that is specifically suited to the 802.11 MAC.

The 802.11e specification adds the Traffic Identifier (TID) concept to 802.11,
which is used to distinguish between different QoS levels. Each station has 16
TIDs assigned (four identifiers for each of the four QoS priority levels), and
packets are commonly mapped to TIDs based on their DiffServ
markings \cite{ietf-802-11}. The TID is attached to every frame when it goes
over the air, and the 802.11n standard specifies that aggregation must be
performed on a per-TID level. Because of this need to do aggregation on a
per-TID basis, it is necessary to be able to separate packets by their assigned
TIDs. Having a separate logical queue for each TID is the natural way to achieve
this, and so this is also the basis for our queueing structure.

To manage the individual per-TID queues, we adapt the FQ-CoDel queue management
scheme, which has been shown to be a best-in-class bufferbloat mitigation
technique \cite{good-bad-wifi,cablelabs,much-ado}. The original FQ-Codel
algorithm is a hybrid fairness queueing and AQM algorithm \cite{fq-codel}. It
functions as a Deficit Round-Robin (DRR) scheduler
\cite{shreedhar_efficient_1996} between flows, hashing packets into queues based
on their transport protocol flows. Each queue has a deficit which controls when
it is allowed to transmit packets: Sending data causes the deficit to drop, and
the round-robin mechanism ensures all back-logged flows recover from negative
deficits at the same rate. The size of the quantum controls the granularity of
the fairness between flows. FQ-CoDel also adds an optimisation for sparse flows
to the basic DRR algorithm. This optimisation allows flows that use less than
their fair share of traffic to gain scheduling priority, reducing the time their
packets spend in the queue. Finally, the CoDel AQM is applied separately to each
queue, in order to keep the latency experienced by each flow under control. For
a full explanation of FQ-CoDel, see \cite{fq-codel}.

\begin{algorithm}[b]
\caption{\small 802.11 queue management algorithm - enqueue.}
\label{alg:fq-codel-enq}
\begin{algorithmic}[1]
\small
\Function{enqueue}{\emph{pkt}, \emph{tid}} \label{ln:beg-enq}
  \If {queue\_limit\_reached()} \Comment{Global limit}
    \State $\textit{drop\_queue} \gets$ \Call{find\_longest\_queue()}{}
    \State \Call{drop}{\textit{drop\_queue.head\_pkt}}
  \EndIf
  \State $\textit{queue} \gets$ \Call{hash}{\emph{pkt}}
  \If {$\textit{queue.tid} \neq$ NULL and $\textit{queue.tid} \neq \textit{tid}$} \label{ln:beg-chk-st}
    \State $\textit{queue} \gets \textit{tid.overflow\_queue}$
  \EndIf \label{ln:end-chk-st}
  \State $\textit{queue.tid} \gets \textit{tid}$ \label{ln:beg-queue-pkt}
  \State \Call{timestamp}{\emph{pkt}} \Comment{Used by CoDel at dequeue} \label{ln:enq-timestamp}
  \State \Call{append}{\emph{pkt}, \emph{queue}} \label{ln:end-queue-pkt}
  \If {\emph{queue} is not active}
    \State \Call{list\_add}{\emph{queue}, \emph{tid.new\_queues}} \label{ln:act-q}
  \EndIf
\EndFunction \label{ln:end-enq}
\end{algorithmic}
\end{algorithm}

FQ-CoDel allocates a number of sub-queues that are used for per-flow scheduling,
and so simply assigning a full instance of FQ-CoDel to each TID is impractical.
Instead, we innovate on the FQ-CoDel design by having it operate on a fixed
total number of queues, and group queues based on which TID they are associated
with. So when a packet is hashed and assigned to a queue, that queue is in turn
assigned to the TID the packet is destined for. In case a hash collision occurs
and the queue is already active and assigned to another TID, the packet is
instead queued to a TID-specific overflow queue. A global queue size limit is
kept, and when this is exceeded, packets are dropped from the globally longest
queue, which prevents a single flow from locking out other flows on overload.
The full enqueue logic is shown in Algorithm \ref{alg:fq-codel-enq}.

The lists of active queues are kept in a per-TID structure, and when a TID needs
to dequeue a packet, the FQ-CoDel scheduler is applied to the TID-specific lists
of active queues. A global limit on the number of queued packets is applied,
with packets being dropped from the (globally) longest queue on overflow. This
is shown in Algorithm \ref{alg:fq-codel-deq}.

\begin{figure}[t]
\centering
\includegraphics[width=\linewidth]{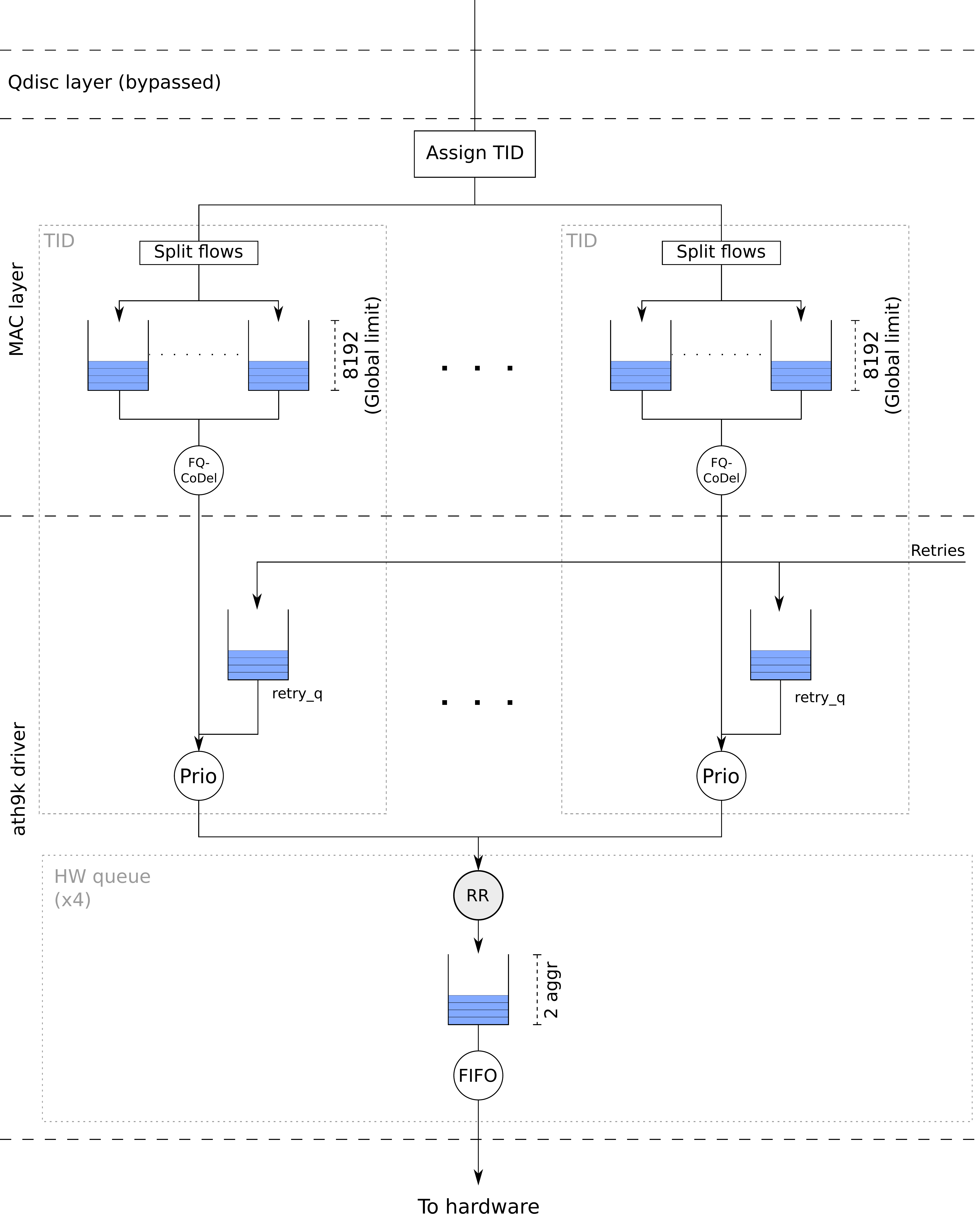}
\caption{\label{fig:linux-queue-post}
Our 802.11-specific queueing structure, as it looks when applied to the Linux WiFi stack.}
\end{figure}

The resulting queueing structure is depicted in Figure \ref{fig:linux-queue-post}. One
additional thing to notice about this structure is that packets can be dequeued
in a different order than they are enqueued (because of the per-flow
scheduling). Thus, any protocol-specific encoding that is sensitive to
reordering (notably 802.11 sequence numbers and encryption IVs) will cause
packets to be discarded by the receiver if they are reordered and thus needs to
be applied on dequeue.

\begin{algorithm}[t]
\caption{\small 802.11 queue management algorithm - dequeue.}
\label{alg:fq-codel-deq}
\begin{algorithmic}[1]
\small
\Function {dequeue}{\emph{tid}} \label{ln:beg-deq}
  \If {\emph{tid.new\_queues} is non-empty} \label{ln:beg-sel-q}
    \State \emph{queue} $\gets$ \Call{list\_first}{\emph{tid.new\_queues}}
  \ElsIf {\emph{tid.old\_queues} is non-empty}
    \State \emph{queue} $\gets$ \Call{list\_first}{\emph{tid.old\_queues}}
  \Else
    \State \Return NULL
  \EndIf \label{ln:end-sel-q}
  \If {$\textit{queue.deficit} \leq 0$} \label{ln:beg-chk-q-def}
    \State $\textit{queue.deficit} \gets \textit{queue.deficit} + \textit{quantum}$
    \State \Call{list\_move}{\emph{queue}, \emph{tid.old\_queues}}
    \State \textbf{restart}
  \EndIf \label{ln:end-chk-q-def}
  \State \emph{pkt} $\gets$ \Call{codel\_dequeue}{\emph{queue}} \label{ln:deq-pkt}
  \If {\emph{pkt} is NULL} \Comment{queue empty} \label{ln:beg-q-empt}
    \If {$\textit{queue} \in \textit{tid.new\_queues}$}
      \State \Call{list\_move}{\emph{queue}, \emph{tid.old\_queues}}
    \Else
      \State \Call{list\_del}{\emph{queue}}
      \State \emph{queue.tid} $\gets$ NULL
    \EndIf
    \State \textbf{restart}
  \EndIf \label{ln:end-q-empt}
  \State \emph{queue.deficit} $\gets$ \emph{queue.deficit} $-$ \emph{pkt.length} \label{ln:dec-q-def}
  \State \Return \emph{pkt} \label{ln:ret-pkt}
\EndFunction \label{ln:end-deq}
\end{algorithmic}
\end{algorithm}

\subsubsection{Dealing with burstiness in the MAC}
\label{sec:bursty-mac}
The WiFi MAC is bursty in nature because stations have to contend for the media
with each other. This means that more queueing is needed than for a less bursty
MAC such as full-duplex switched Ethernet. This is especially true after the
introduction of aggregation in 802.11n and later standards. The CoDel AQM
employed on each queue can become too aggressive when applied to WiFi
traffic. In addition, as others have shown before \cite{good-bad-wifi,much-ado},
CoDel needs tuning for very low bandwidths, or it will drop too aggressively. In
WiFi, transmission rates can vary over a large range and different stations can
have different rates. This is important to take into account when tuning the
behaviour of an AQM. To deal with this, we assign CoDel parameters per station
instead of globally. The parameters are assigned per station rather than per TID
because the link characteristics varies with the physical medium, which is a
property of each station, but identical between TIDs assigned to the same
station.

We use a simple threshold combined with an estimate of the station's current
throughput, obtained from the rate selection algorithm, changing CoDel's
\emph{target} to 50 ms and \emph{interval} to 300 ms when the expected rate drops
below 12 Mbps. We apply hysteresis so the values are not changed more than
once every two seconds. We have found this simple mechanism avoids the worst
starvation in our setup, and so leave refinements to future work.

\subsection{Airtime fairness scheduling}
\label{sec:airtime-scheduler}
Given the above queueing structure, it achieving airtime fairness becomes a
matter of measuring the airtime used by each station, and appropriately
scheduling the order in which stations are served. For each packet sent or
received, the packet duration can either be extracted directly from a hardware
register, or it can be calculated from the packet length and the rate at which
it was sent (including any retries). Each packet's duration is subtracted from
an airtime deficit that is kept for each QoS level for each station (so four
deficits per station, corresponding to the VO,VI,BE and BK 802.11 precedence
levels). This deficit is then used by a deficit scheduler modelled after
FQ-CoDel to decide which station receives the next transmission.

The resulting airtime fairness scheduler is shown in
Algorithm \ref{alg:airtime-scheduler}. It is similar to the the FQ-CoDel dequeue
algorithm, with stations taking the place of flows, and the deficit being
accounted in microseconds instead of bytes. The two main differences are (1)
that the scheduler function loops until the hardware queue becomes full (at two
queued aggregates), rather than just dequeueing a single packet; and (2) that
when a station is chosen to be scheduled, it gets to build a full aggregate
rather than a single packet.

Compared to the closest previously proposed
solution \cite{garroppo_providing_2007}, our scheme has several advantages:

\begin{enumerate}
\item We lower implementation complexity by leveraging already existing information
on per-aggregate transmission rates and time, and by using a per-station
deficit instead of token buckets, which means that no token bucket accounting
needs to be performed at TX and RX completion time.

\item \cite{garroppo_providing_2007} measures time from an aggregate is submitted to
the hardware until it is sent, which risks including time spent waiting for
other stations to transmit. We increase accuracy by only measuring the actual
time spent transmitting, and by also accounting the airtime from received
frames to each station's deficit.

\item We improve on the basic scheduler design by adding an optimisation for sparse
stations, analogous to FQ-CoDel's sparse flow optimisation. This improves
latency for stations that only transmit occasionally, by giving them
temporary priority for one round of scheduling (but not more; we apply the
same protection against gaming this mechanism that FQ-CoDel does to its
sparse flow mechanism \cite{fq-codel}).
\end{enumerate}

\begin{algorithm}
\caption{\small Airtime fairness scheduler. The schedule function is called when new packets arrive and after transmission completes.}
\label{alg:airtime-scheduler}
\begin{algorithmic}[1]
\small
\Function {schedule}{} \label{ln:beg-schedule}
  \While {hardware queue is not full}
    \If {\emph{new\_stations} is non-empty}
      \State \emph{station} $\gets$ \Call{list\_first}{\emph{new\_stations}}
    \ElsIf {\emph{old\_stations} is non-empty}
      \State \emph{station} $\gets$ \Call{list\_first}{\emph{old\_stations}}
    \Else
      \State \Return
    \EndIf
    \If {$\textit{station.deficit} \leq 0$}
      \State $\textit{station.deficit} \gets $ {\small$\textit{station.deficit} + \textit{quantum}$ }
      \State \Call{list\_move}{\emph{station}, \emph{old\_stations}}
      \State \textbf{restart}
    \EndIf
    \If {\emph{station}'s queue is empty}
      \If {$\textit{station} \in \textit{new\_stations}$}
        \State \Call{list\_move}{\emph{station}, \emph{old\_stations}}
      \Else
        \State \Call{list\_del}{\emph{station}}
      \EndIf
      \State \textbf{restart}
    \EndIf
    \State \Call{build\_aggregate}{\emph{station}} \label{ln:q-aggr}
  \EndWhile
\EndFunction \label{ln:end-schedule}
\end{algorithmic}
\end{algorithm}

\subsection{Implementation}
\label{sec:org02727bc}
We have implemented our proposed queueing scheme in the Linux kernel, modifying
the mac80211 subsystem to include the queueing structure itself, and modifying
the ath9k and ath10k drivers for Qualcomm Atheros 802.11n and 802.11ac chipsets
to use the new queueing structure. The airtime fairness scheduler implementation
is limited to the ath9k driver, as the ath10k driver lacks the required
scheduling hooks.

Our modifications have been accepted into the mainline Linux kernel, different
parts going into kernel releases 4.8 through 4.11. The implementation is
available online, as well as details about our test environment and the full
evaluation dataset.\footnote{See \url{http://www.cs.kau.se/tohojo/airtime-fairness/} for the online appendix
that contains additional material, as well as the full experimental dataset and
links to the relevant Linux kernel patches.}
\section{Evaluation}
\label{sec:evaluation}
We evaluate our modifications in a testbed setup consisting of five PCs: Three
wireless clients, an access point, and a server located one Gigabit Ethernet hop
from the access point, which serves as source and sink for the test flows. All
the wireless nodes are regular x86 PCs equipped with PCI-Express Qualcomm
Atheros AR9580 adapters (which use the ath9k driver). Two of the test clients
are placed in close proximity to the access point (and are referred to as fast
nodes), while the last (referred to as the slow node) is placed further away and
configured to only support the MCS0 rate, giving a maximum throughput to that
station of 7.2 Mbps at the PHY layer. A fourth virtual station is added as an
additional fast node to evaluate the sparse station optimisation (see
Section \ref{sec:sparse} below). All tests are run in HT20 mode on an otherwise
unused channel in the 5Ghz band. We use 30 test repetitions of 30 seconds each
unless noted otherwise.

The wireless nodes run an unmodified Ubuntu 16.04 distribution. The access point
has had its kernel replaced with a version 4.6 kernel from kernel.org on top of
which we apply our modifications. We run all experiments with four queue
management schemes, as follows:

\begin{itemize}
\item \textbf{FIFO}: The default 4.6 kernel from kernel.org modified only to collect
the airtime used by stations, running with the default PFIFO queueing
discipline installed on the wireless interface.
\item \textbf{FQ-CoDel}: As above, but using the FQ-CoDel qdisc on the wireless interface.
\item \textbf{FQ-MAC}: Kernel patched to include the FQ-CoDel based intermediate queues in
the MAC layer (patching the mac80211 subsystem and the ath9k driver).
\item \textbf{Airtime fair FQ}: As FQ-MAC, but additionally modified to include our airtime
fairness scheduler in the ath9k driver.
\end{itemize}

Our evaluation is split into two parts. First, we validate the effects of the
modifications in simple scenarios using synthetic benchmark traffic. Second, we
evaluate the effect of our modifications on two application traffic scenarios,
to verify that they provide a real-world benefit.

\subsection{Validation of effects}
\label{sec:orgea833e2}
In this section we present the evaluation of our modifications in simple
synthetic scenarios designed to validate the correct functioning of the
algorithms and to demonstrate various aspects of their performance.

\subsubsection{Latency reductions}
\label{sec:orgdd60700}
\begin{figure}[htbp]
\centering
\includegraphics[width=\linewidth]{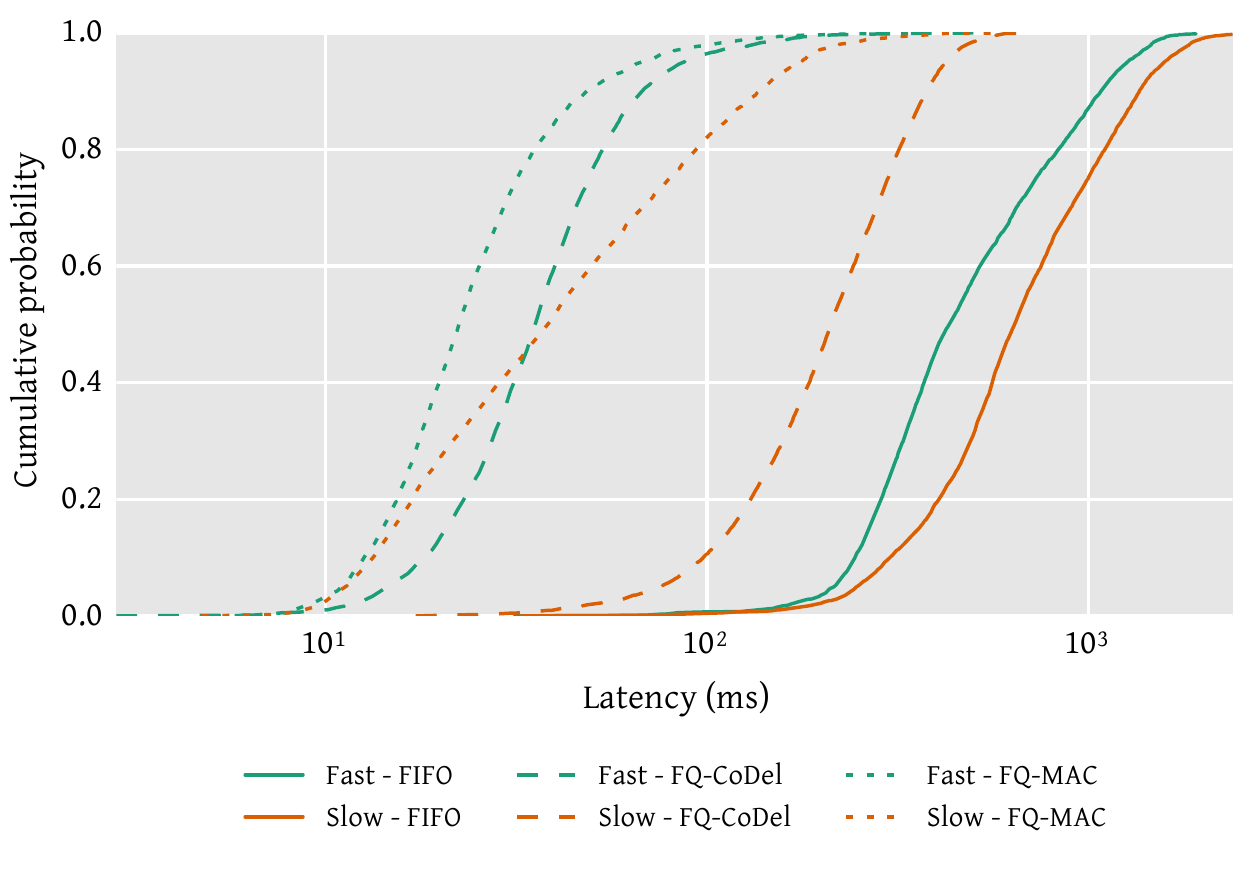}
\caption{\label{fig:tcp-up-latency}
Latency (ICMP ping) with simultaneous TCP download traffic.}
\end{figure}

Figure \ref{fig:tcp-up-latency} is the full set of results for our ICMP latency
measurements with simultaneous TCP download traffic (of which a subset was shown
earlier in Figure \ref{fig:tcp-up-latency-partial}). Here, the FIFO case shows several
hundred milliseconds of latency when the link is saturated by a TCP download.
FQ-CoDel alleviates this somewhat, but the slow station still sees latencies of
more than 200 ms in the median, and the fast stations around 35 ms. With the
FQ-MAC queue restructuring, this is reduced so that the slow station now has the
same median latency as the fast one does in the FQ-CoDel case, while the fast
stations get their latency reduced by another 45\%. The airtime scheduler doesn't
improve further upon this, other than to alter the shape of the distribution
slightly for the slow station (but retaining the same median). For this reason,
we have omitted it from the figure to make it more readable.

For simultaneous upload and download the effect is similar, except that in this
case the airtime scheduler slightly worsens the latency to the slow station,
because it is scheduled less often to compensate for its increased airtime usage
in the upstream direction. The graph of this case can be found in the online
appendix.

\subsubsection{Airtime usage}
\label{sec:org4903cc4}
\begin{figure}[htbp]
\centering
\includegraphics[width=\linewidth]{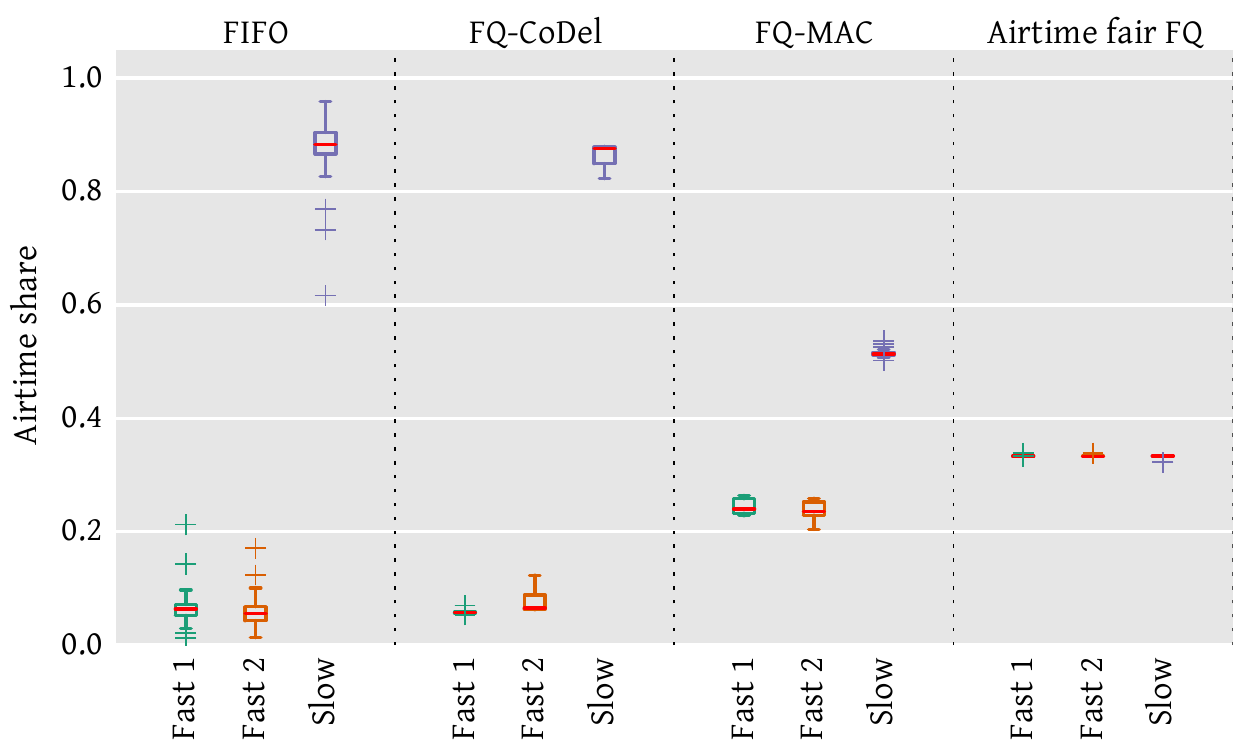}
\caption{\label{fig:airtime-udp}
Airtime usage for one-way UDP traffic. Each column shows the relative airtime usage of one of the three stations, with the four sections corresponding to the four queue management schemes.}
\end{figure}

Figure \ref{fig:airtime-udp} shows the airtime usage of the three active stations for
one-way UDP traffic going to the stations. There is no reverse traffic and no
contention between stations, since only the access point is transmitting data.
This is the simplest case to reason about and measure, and it clearly shows the
performance anomaly is present in the current Linux kernel (left half of the
figure): The third station (which transmits at the lowest rate) takes up around
80 \% of the available airtime, while the two other stations share the remaining
20 \%.

The differences between the first two columns and the third column are due to
changes in aggregation caused by the change to the queueing structure. In the
FIFO and FQ-CoDel cases, there is a single FIFO queue with no mechanism to
ensure fair sharing of that queue space between stations. This means that
because the slow station has a lower egress rate, it will build more queue until
it occupies the entire queueing space. This means that there are not enough
packets queued to build sufficiently large aggregates for the fast stations to
use the airtime effectively. The FQ-MAC queueing scheme drops packets from the
largest queue on overflow, which ensures that the available queueing space is
shared between stations, which improves aggregation for the fast stations and
thus changes the airtime shares. Referring back to
Table \ref{tbl:airtime-rates}, the values correspond well to those predicted by
the analytical model. The fourth column shows the airtime fairness scheduler
operating correctly: Each station receives exactly the same amount of airtime in
this simple one-way test.

\begin{figure}[htbp]
\centering
\includegraphics[width=\linewidth]{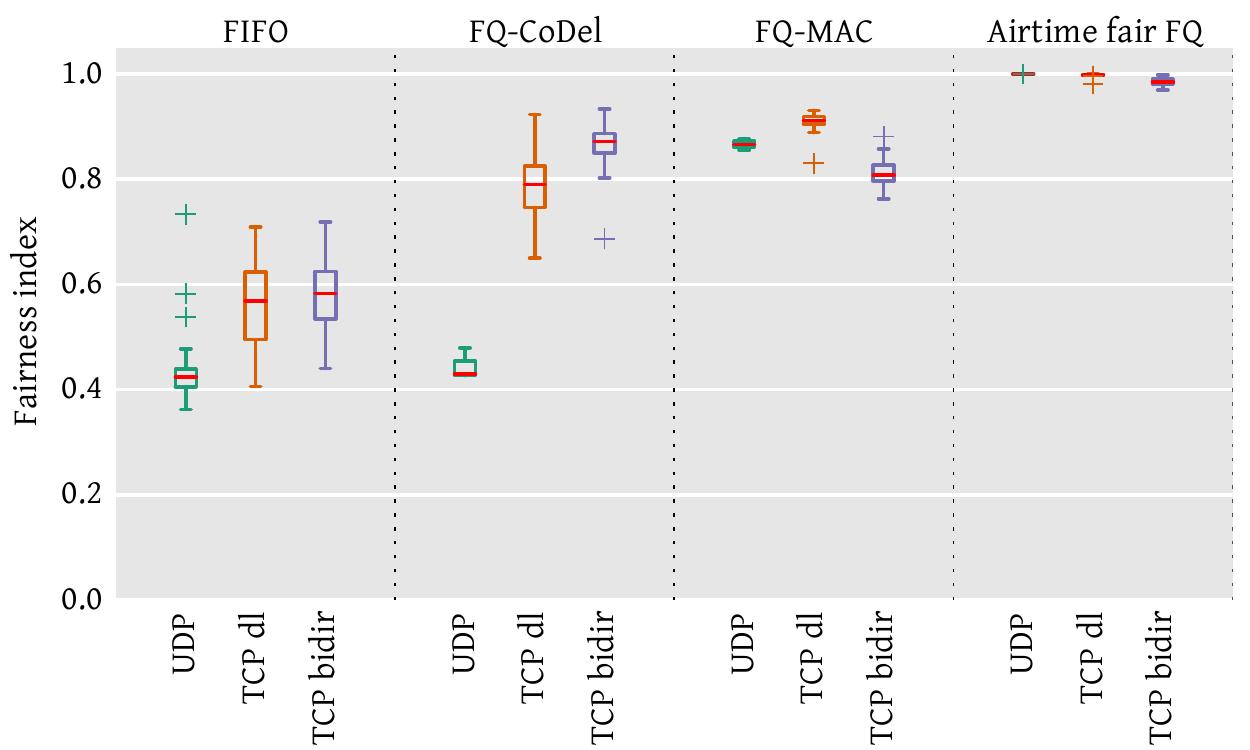}
\caption{\label{fig:airtime-fairness}
Jain's fairness index computed over the airtime fairness usage between the three stations, for UDP traffic, TCP download, and simultaneous TCP upload and download traffic.}
\end{figure}

Going beyond the simple UDP case, Figure \ref{fig:airtime-fairness} shows Jain's
fairness index for the airtime of the four different schemes for UDP (for
comparison) and both unidirectional (to the clients) and bidirectional
(simultaneous up and down) TCP traffic. The same general pattern is seen with
TCP as with UDP traffic: The performance anomaly is clear for the FIFO case, but
somewhat lessened for the FQ-CoDel and FQ-MAC cases. The airtime fairness
scheduler achieves close to perfect sharing of airtime in the case of
uni-directional traffic, with a slight dip for bidirectional traffic. The latter
is because the scheduler only exerts indirect control over the traffic sent from
the clients, and so cannot enforce perfect fairness as with the other traffic
types. However, because airtime is also accounted for received packets, the
scheduler can partially compensate, which is why the difference between the
unidirectional and bidirectional cases is not larger than it is.

\subsubsection{Effects on throughput}
\label{sec:org7f142ff}
\begin{figure}[htbp]
\centering
\includegraphics[width=\linewidth]{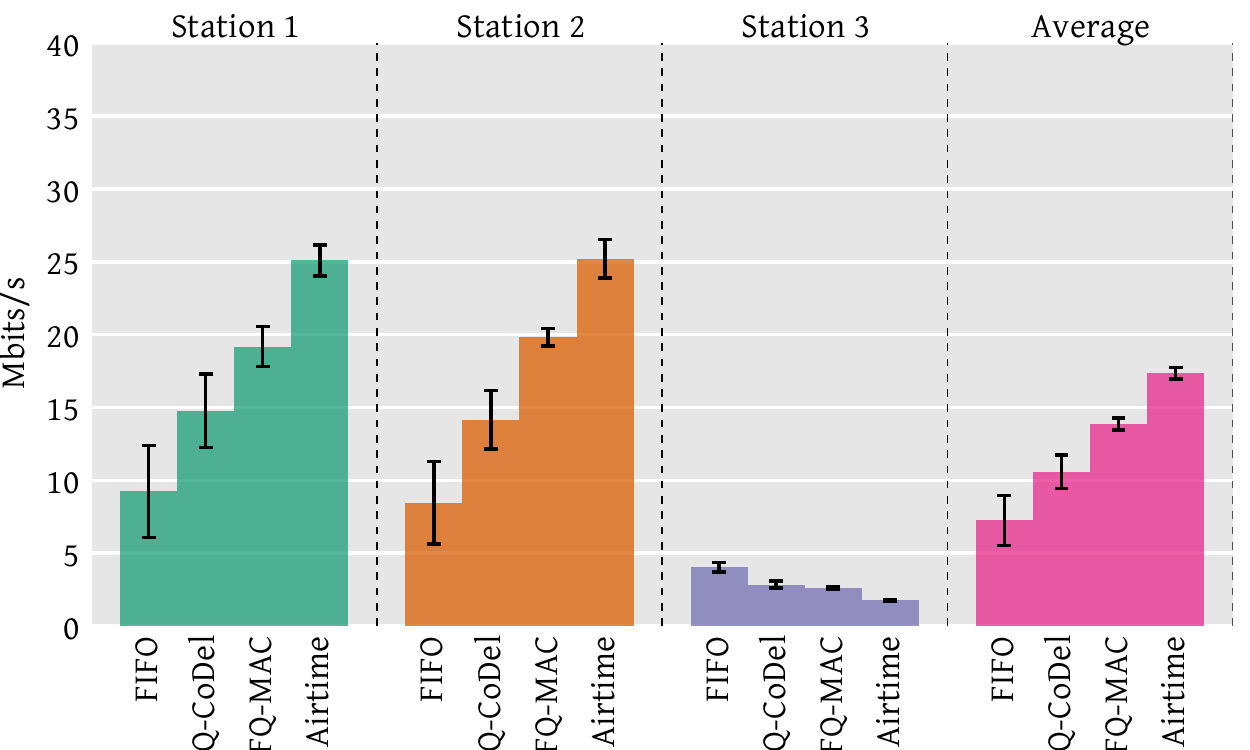}
\caption{\label{fig:throughput-tcp-up}
Throughput for TCP download traffic (to clients).}
\end{figure}

As was already shown in Table \ref{tbl:airtime-rates}, fixing the performance
anomaly improves the efficiency of the network for unidirectional UDP traffic.
Figure \ref{fig:throughput-tcp-up} shows the throughput for downstream TCP traffic.
For this case, the fast stations increase their throughput as fairness goes up,
and the slow station decreases its throughput. The total effect is a net
increase in throughput. The increase from the FIFO case to FQ-CoDel and FQ-MAC
is due to better aggregation for the fast stations. This was observed for UDP as
well in the case of FQ-MAC, but for FQ-CoDel the slow station would occupy all
the queue space in the driver, preventing the fast station from achieving full
aggregation. With the TCP feedback loop in place, this lock-out behaviour is
lessened, and so fast stations increase their throughput.

When traffic is flowing in both directions simultaneously, the pattern is
similar, but with a slightly higher variance. The graph for the bidirectional
case can be found in the online appendix.

\subsubsection{The sparse station optimisation}
\label{sec:sparse}
\begin{figure}[htbp]
\centering
\includegraphics[width=\linewidth]{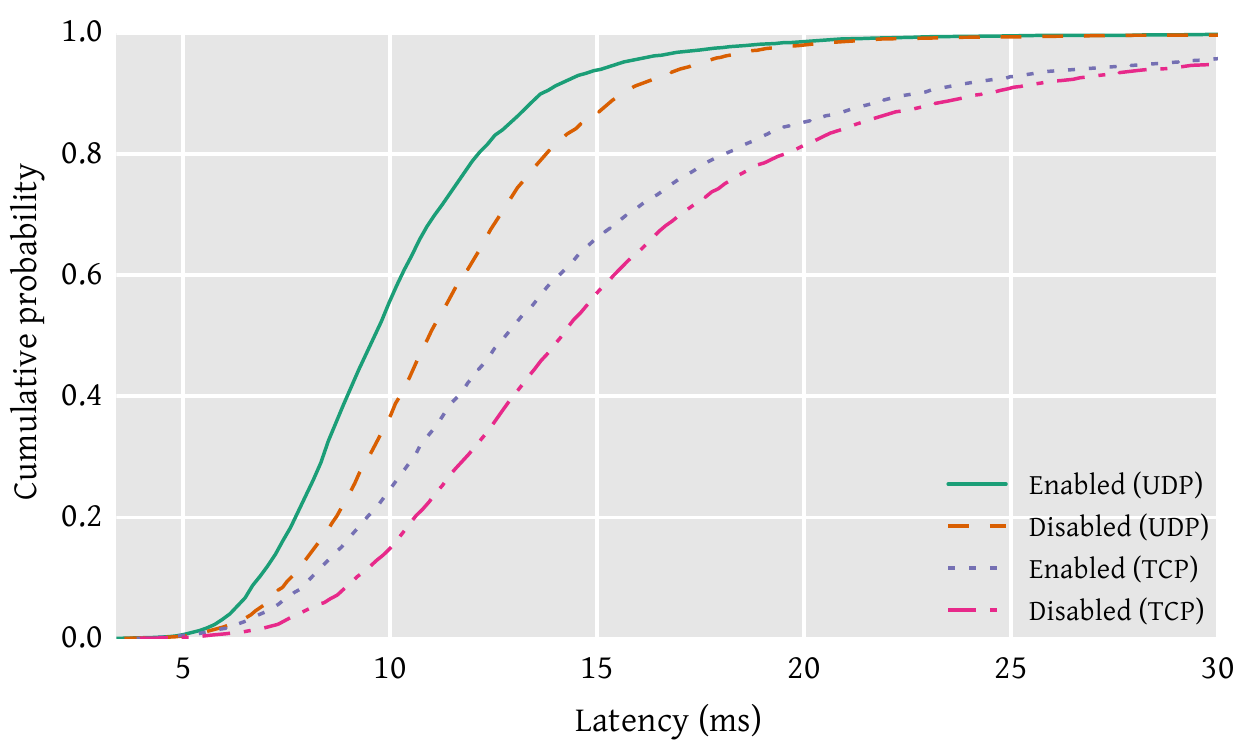}
\caption{\label{fig:sparse-stations}
The effects of the sparse station optimisation.}
\end{figure}

To evaluate the impact of the sparse station optimisation, we add a fourth
station to our experiments which receives only a ping flow, but no other
traffic, while the other stations receive bulk traffic as above. We measure the
latency to this extra station both with and without the sparse station
optimisation. The results of this are shown in Figure \ref{fig:sparse-stations}, for
both UDP and TCP download traffic. In both cases, a small, but consistent,
improvement is visible: the round-trip latency to the fourth station is reduced
by \(10-15\%\) (in the median) when the optimisation is in place.

\subsubsection{Scaling to more stations}
\label{sec:org60227d2}
While the evaluations presented in the previous sections have shown that our
modifications work as planned, and that they provide a substantial benefit in a
variety of scenarios, one question is left unanswered: Does the solution scale
to more stations? To answer this, we arranged for an independent third party to
repeat a subset of our tests in their testbed, which features an access point
and 30 clients. The nodes are all embedded wireless devices from a commercial
vendor that bases its products on the OpenWrt/LEDE open-source router platform,
running a LEDE firmware development snapshot from November 2016.

In this setup, one of the 30 clients is artificially limited to only transmit at
the lowest possible rate (1 Mbps, i.e. disabling HT mode), while the others are
configured to select their rate in the usual way, on a HT20 channel in the 2.4
Ghz band. One of the 29 "fast" clients only receives ping traffic, leaving 28
stations to contend with the slow 1 Mbps station for airtime and bandwidth.

\begin{figure}[htbp]
\centering
\includegraphics[width=\linewidth]{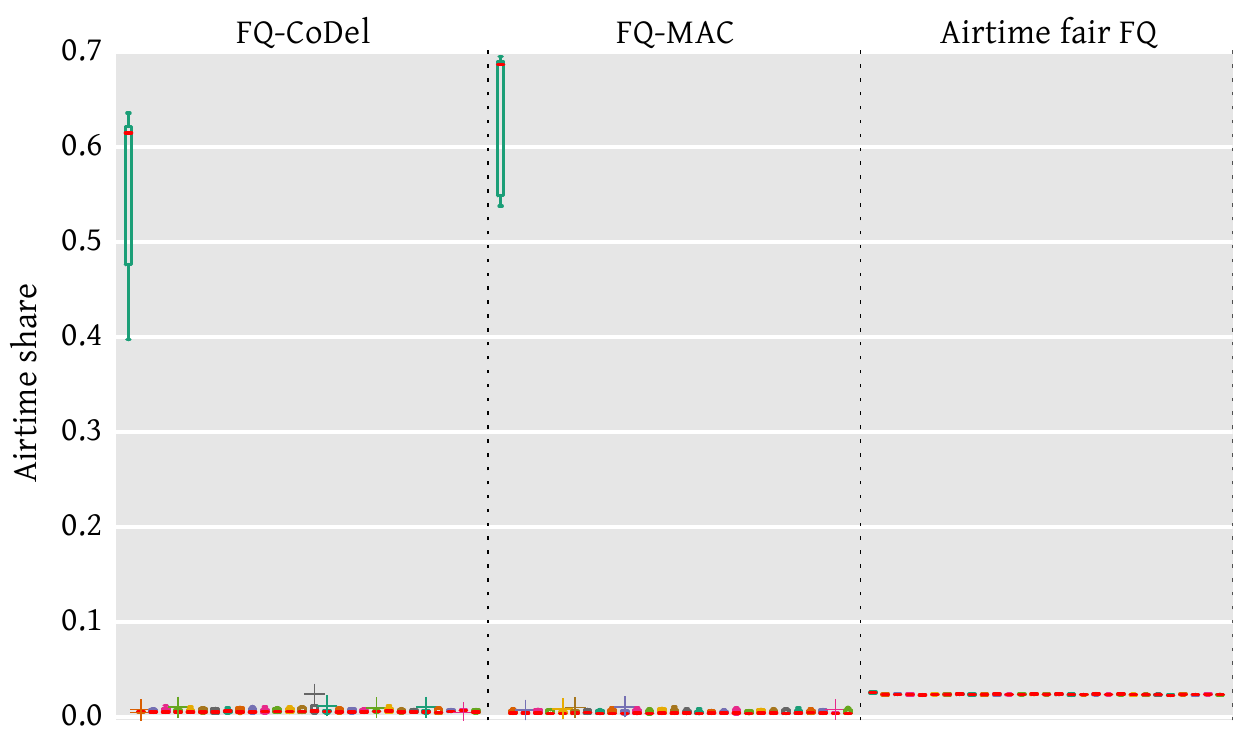}
\caption{\label{fig:30-stations-tcp-airtime}
Airtime share between stations in the 30 stations TCP test.}
\end{figure}

\begin{figure}[htbp]
\centering
\includegraphics[width=\linewidth]{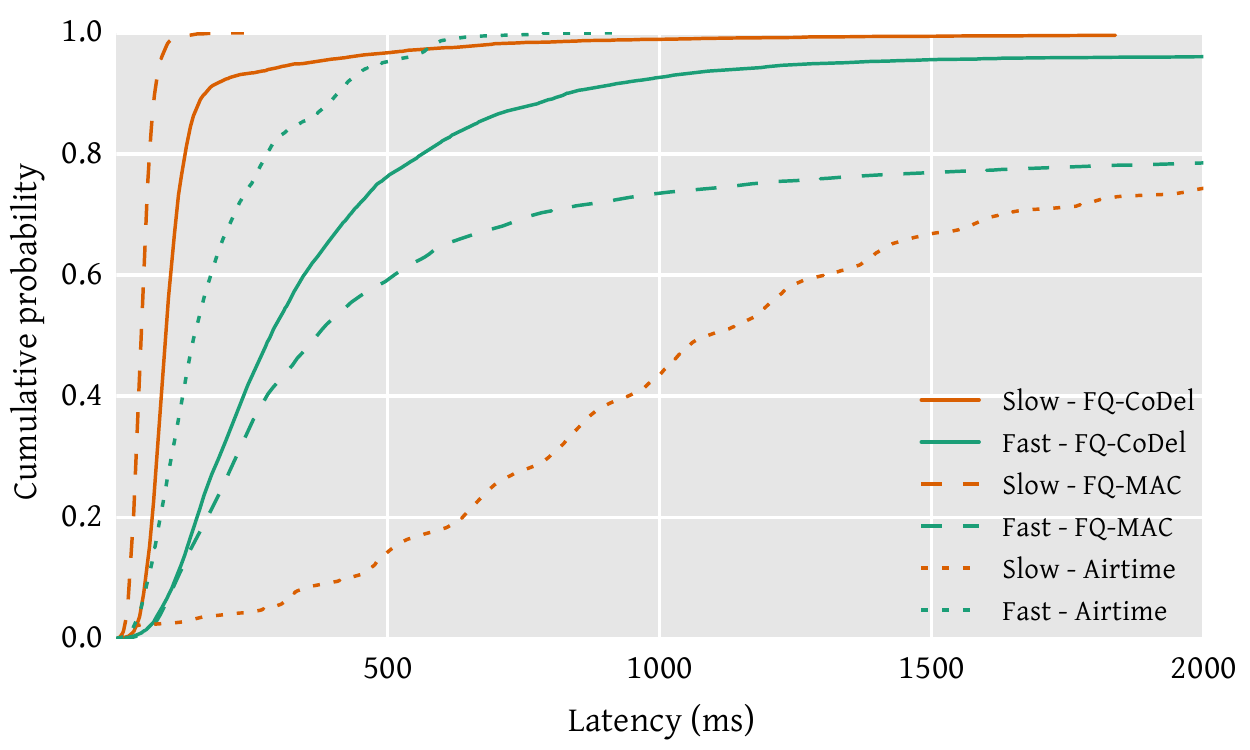}
\caption{\label{fig:30-stations-tcp-latency}
Latency for the 30 stations TCP test.}
\end{figure}

In this environment, our downstream TCP experiment presented above was repeated,
with the difference that each test was run for five minutes, but with only five
repetitions, and without the FIFO test case. A subset of these results are shown
in figures \ref{fig:30-stations-tcp-airtime} and \ref{fig:30-stations-tcp-latency}, which
show airtime usage and latency respectively. From this experiment, we make
several observations:

\begin{enumerate}
\item When the slow station is at this very low rate, it manages to grab around two
thirds of the available airtime, even with 28 other stations to compete with.
However, our airtime fairness scheduler manages to achieve completely fair
sharing of airtime between all 29 stations.

\item Total throughput goes from a mean of \(3.3\) Mbps for the FQ-CoDel case to
\(17.7\) Mbps with the airtime scheduler. That is, the relative gain of
throughput with airtime fairness is 5.4x in this scenario.\footnote{Not shown in the figures. See the online appendix.}

\item As can be expected, with the airtime fairness scheduler, the latency to the
fast stations is improved with the increased throughput, but the latency to
the slow station increases by an order of magnitude in the median. Overall,
the \emph{average} latency (to all stations) is improved by a factor of two.

\item With 30 stations, we see the sparse station optimisation being even more
effective; in this scenario it reduces latency to the sparse station by a
factor of two.\footnotemark[4]
\end{enumerate}

Finally, we verify the in-kernel airtime measurement against a tool developed by
the same third party that measures airtime from captures taken with a monitor
device. We find that the two types of measurements agree to within 1.5\%, on
average.

\subsection{Effects on real-world application performance}
\label{sec:orgdd1e253}
In the previous section we evaluated our solution in a number of scenarios
that verify its correct functioning and quantify its benefits. In this section
we expand on that validation by examining how our modifications affect
performance of two important real-world applications: VoIP and web browsing.

\subsubsection{VoIP}
\label{sec:orgfc42688}
VoIP is an important latency-sensitive application which it is desirable to have
working well over WiFi, since that gives mobile handsets the flexibility of
switching between WiFi and cellular data as signal conditions change. To
evaluate our modifications in the context of VoIP traffic, we measure VoIP
performance when mixed with bulk traffic. As in Section \ref{sec:sparse} we
include a virtual station as another fast station, and so these scenarios have
three fast stations. Due to space constraints, we only include the case where
the slow station receives both VoIP traffic \emph{and} bulk traffic, while the fast
stations receive bulk traffic. The other cases show similar relative performance
between the different queue management schemes.

The QoS markings specified in the 802.11e standard can be used to improve the
performance of VoIP traffic, and so we include this aspect in our evaluation.
802.11e specifies four different QoS levels, of which voice (VO) has the highest
priority. Packets transmitted with this QoS marking gets both queueing priority
and a shorter contention window, but cannot be aggregated. This difference can
dramatically reduce the latency of the traffic, at a cost in throughput because
of the lack of aggregation. We repeat the voice experiments in two variants: One
where the VoIP packets are sent as best effort traffic, and one where they are
put into the high-priority VO queue. We also repeat the tests with a baseline
one-way delay of both 5 ms and 50 ms.

To serve as a metric of voice quality, we calculate an estimate of the Mean
Opinion Score (MOS) of the VoIP flow according to the E-model specified in the
ITU-T G.107 recommendation \cite{_e-model:_2015}. This model can predict the MOS
from a range of parameters, including the network conditions. We fix all audio
and codec related parameters to their default values and calculate the MOS
estimate based on the measured delay, jitter and packet loss. The model gives
MOS values in the range from \(1-4.5\).

Table \ref{tbl:voip-mos} shows the results. As expected, the FIFO and FQ-CoDel
cases have low MOS values when the voice traffic is marked as BE, and higher
values when using the VO queue. However, both the FQ-MAC and airtime fairness
schemes achieve \emph{better} MOS values with best-effort traffic than the unmodified
kernel does with VO-marked traffic. In the FQ-MAC and airtime cases, using the
VO queue still gives a slightly better MOS score than using the BE queue does;
but the difference is less than half a percent. This is an important
improvement, because it means that with our modifications, applications can rely
on excellent real-time performance even when not in control of the DiffServ
markings of their traffic

\begin{table}[htb]
\centering
\small
\begin{tabular}{lcp{1pt}ccp{1pt}cc}
\toprule
& && \multicolumn{2}{c}{5 ms} && \multicolumn{2}{c}{50 ms} \\
 \cmidrule{4-5}  \cmidrule{7-8}
 & QoS && MOS & Thrp &&  MOS & Thrp\\
\midrule
\multirow{2}{*}{FIFO} & VO && 4.17 & 27.5 && 4.13 & 21.6\\
 & BE && 1.00 & 28.3 && 1.00 & 22.0\\
\midrule
\multirow{2}{*}{FQ-CoDel} & VO && 4.17 & 25.5 && 4.08 & 15.2\\
 & BE && 1.24 & 23.6 && 1.21 & 15.1\\
\midrule
\multirow{2}{*}{FQ-MAC} & VO && 4.41 & 39.1 && 4.38 & 28.5\\
 & BE && 4.39 & 43.8 && 4.37 & 34.0\\
\midrule
\multirow{2}{*}{Airtime} & VO && 4.41 & 56.3 && 4.38 & 49.8\\
 & BE && 4.39 & 57.0 && 4.37 & 49.7\\
\bottomrule
\end{tabular}
\caption{\label{tbl:voip-mos}
MOS values and total throughput when using different QoS markings for VoIP traffic.
Data for 5 ms and 50 ms baseline one-way delay.}
\end{table}

\subsubsection{Web}
\label{sec:orgbf421cc}
Another important real-world application is web traffic. To investigate the
impact of our modifications on this, we measure page load time (PLT) with
emulated web traffic using a web client based on the cURL library. The client
mimics the common web browser behaviour of fetching multiple requests in
parallel over four different TCP connections. We simply measure the total time
to fetch a web site and all its attached resources (including the initial DNS
lookup) for two different pages: A small page (56 KB data in three requests) and
a large page (3 MB data in 110 requests). We run the experiments in two
scenarios: One where a fast station fetches the web sites while the slow station
runs a simultaneous bulk transfer, to emulate the impact of a slow station on
the browsing performance of other users on the network. And another scenario
where the slow station fetches the web sites while the fast stations run
simultaneous bulk transfers, to emulate the browsing performance of a slow
station on a busy network.

\begin{figure}[h]
\centering
\includegraphics[width=\linewidth]{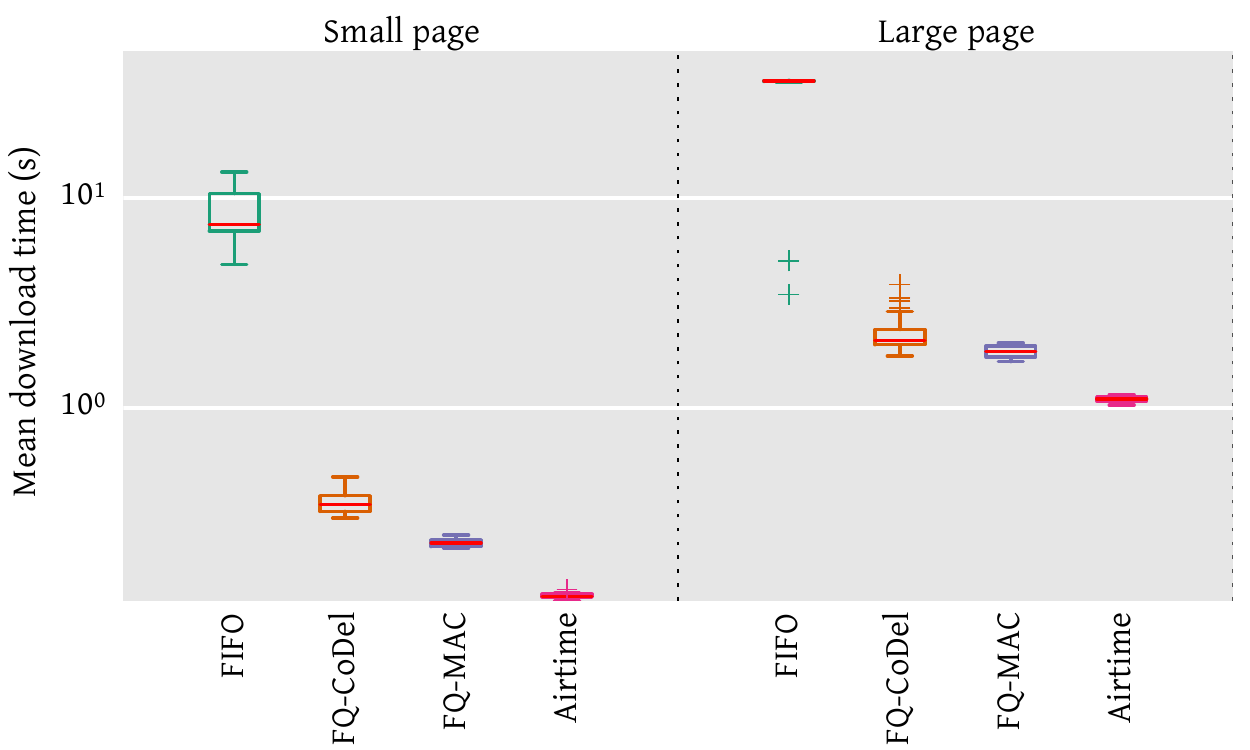}
\caption{\label{fig:http-fast}
HTTP page fetch times for a fast station (while the slow station runs a bulk transfer).  Note the log scale - the fetch time for the large page is 35 \emph{seconds} for the FIFO case.}
\end{figure}

The results for the fast station are seen in Figure \ref{fig:http-fast}: Fetch times
decrease from the FIFO case as the slowest to the airtime fair FQ case as the
fastest. In particular, there is a an order-of-magnitude improvement when going
from FIFO to FQ-CoDel, which we attribute to the corresponding significant
reduction in latency seen earlier.

When the slow station is fetching the web page, adding airtime fairness
increases page load time by \(5-10\%\). This is as expected since in this case the
slow station is being throttled. The graph for this can be found in the online
appendix.

\subsection{Summary}
\label{sec:orgbf02056}
Our evaluation shows that our modifications achieve their design goal: We
eliminate the performance anomaly and achieve close to perfect airtime fairness
even when station rates vary, and our solution scales successfully as more
clients are added. We improve total throughput by up to a factor of five and
reduce latency under load by up to an order of magnitude. We also achieve close
to perfect airtime fairness in a scenario where traffic is mixed between
upstream and downstream flows from the different stations. Finally, the
optimisation that prioritises sparse stations achieves a consistent improvement
in latency to stations that only receive a small amount of traffic.

In addition, we show that our modifications give significant performance
increases for two important real-world applications: VoIP and web traffic. In
the case of VoIP, we manage to achieve better performance with best effort
traffic than was achievable with traffic marked as Voice according to the
802.11e QoS standard. With web traffic we achieve significant reductions in
the page load time for both large and small web sites.
\section{Related work}
\label{sec:related-work}
There have been several previous studies on bufferbloat and its mitigations
(e.g. \cite{cablelabs,much-ado}), but only a few that deal with the problem in a
WiFi-specific context: \cite{good-bad-wifi} and \cite{much-ado} both feature a
WiFi component in a larger evaluation of bufferbloat mitigation techniques and
show that while these techniques can help on a WiFi link, the lower-level
queueing in the WiFi stack prevents a full solution of the problem in this
space. \cite{showail_buffer_2016} draws similar conclusions while looking at
buffer sizing (but only mentions AQM-based solutions briefly). Finally,
\cite{cai_wireless_2007} touches upon congestion at the WiFi hop and uses
different queueing schemes to address it, but in the context of a centralised
solution that also seek to control fairness in the whole network. None of these
works actually provide a solution for bufferbloat at the WiFi link itself, as we
present here.

Several different schemes to achieve fairness based on modifying the contention
behaviour of nodes are presented in
\cite{jiang_proportional_2005,lin_achieving_2011,sanabria-russo_future_2013,joshi_airtime_2008,heusse_idle_2005,yazici_running_2013}.
\cite{jiang_proportional_2005} and \cite{lin_achieving_2011} both propose
schemes that use either the 802.11e TXOP feature to allocate equal-length to all
stations, or scaling of the contention window by the inverse of the transmission
rate to achieve fairness. \cite{joshi_airtime_2008} develops an analytical model
to predict the values to use for a similar scaling behaviour, which is also
verified in simulation. \cite{sanabria-russo_future_2013} presents a modified
contention behaviour that can lower the number of collisions experienced, but
they do not verify the effect of their schemes on airtime fairness.
\cite{heusse_idle_2005} proposes a modification to the DCF based on sensing the
idle time of the channel scaling CW\(_{\text{min}}\) with the rate to achieve fairness.
Finally, \cite{yazici_running_2013} proposes a scheme for airtime fairness that
runs several virtual DCF instances per node, scaling the number of instances
with the rate and channel properties.

Achieving fairness by varying the transmission size is addressed in
\cite{dunn_practical_2004,razafindralambo_dynamic_2006,kim_adaptive_2015}. The
former two predate the aggregation features of 802.11n and so
\cite{dunn_practical_2004} proposes to scale the packet size downwards by
varying the MTU with the transmission rate. \cite{razafindralambo_dynamic_2006}
goes the other way and proposes a scheme where a station will burst packets to
match the total transmission length of the previous station that was heard on
the network. Finally, \cite{kim_adaptive_2015} uses the two-level aggregation
feature of 802.11n and proposes a scheme to dynamically select the optimal
aggregation size so all transmissions take up the same amount of time.

Turning to schedulers, \cite{gomez_efficient_2011} and
\cite{garroppo_providing_2007} both propose schedulers which work at the access
point to achieve airtime fairness, the former estimating the packet transmission
time from channel characteristics, and the latter measuring it after
transmission has occurred. \cite{riggio_airtime_2008} proposes a solution for
wireless mesh networks, which leverages routing metrics to schedule links in a
way that ensures fairness. Finally, \cite{kliazovich_queue_2011} proposes a
scheme to scale the queue space at the access point based on the BDP of the
flows going through the access point. Our solution is closest to
\cite{garroppo_providing_2007}, but we improve upon it by increasing accuracy
and reducing implementation difficulty, while adding new features to the
algorithm, as was described in Section \ref{sec:airtime-scheduler}.

A few proposals fall outside the categories above:
\cite{kashibuchi_channel_2010} proposes a TCP congestion control algorithm that
uses information about the wireless conditions to cap the TCP window size of
clients to achieve fairness. Finally, there are schemes that sidestep the
fairness problems of the 802.11 MAC and instead replace it entirely with TDMA
scheduling. \cite{ben_salem_fair_2005} proposes a scheme for TDMA scheduling in
a mesh network that ensures fair bandwidth allocation to all connecting clients,
and \cite{torfs_tdma_2015} implements a TDMA transmission scheme for
infrastructure WiFi networks.

\section{Conclusion}
\label{sec:conclusion}
We have developed a novel two-part solution to two large performance problems
affecting WiFi: Bufferbloat and the 802.11 performance anomaly. The solution
consists of a new integrated queueing scheme tailored specifically to
eliminating bufferbloat in WiFi, which reduces latency under load by an order of
magnitude. Leveraging the queueing structure, we have developed a deficit-based
airtime fairness scheduler that works at the access point with no client
modifications, and achieves close to perfect fairness in all the evaluated
scenarios, increasing total throughput by up to a factor of 5.

Our solution reduces implementation complexity and increases accuracy compared
to previous work, and has been accepted into the mainline Linux kernel, making
it deployable on billions of Linux-based devices.

\section{Acknowledgements}
We would like to thank Sven Eckelmann and Simon Wunderlich for their work on
independently verifying our implementation. Their work was funded by Open Mesh
Inc, who also supplied their test hardware. We would also like to thank Felix
Fietkau, Tim Shepard, Eric Dumazet, Johannes Berg, and the numerous other
contributors to the Make-Wifi-Fast and LEDE projects for their insights, review
and contributions to many different iterations of the implementation.

Portions of this work were funded by Google Fiber and by the Comcast Innovation
Fund.

\section{References}
\label{sec:org5674c9e}

\printbibliography[heading=none]
\end{document}